\documentclass[a4paper,11pt]{article}
\usepackage{pos}
\usepackage{sidecap}
\usepackage{wrapfig}

\makeatletter
\g@addto@macro\bfseries{\boldmath}
\makeatother

\title{Status of two-baryon scattering in lattice QCD}

\author*[a]{Jeremy R.\ Green}

\affiliation[a]{Deutsches Elektronen-Synchrotron DESY,\\
  Platanenallee 6, 15738 Zeuthen, Germany}

\emailAdd{jeremy.green@desy.de}

\abstract{In these proceedings, I will review lattice QCD calculations
  of baryon-baryon scattering, their methods, and their challenges. In
  the last few years, there has been a new generation of calculations
  with increased focus on controlling systematic
  uncertainties. Contrary to the findings of earlier exploratory
  calculations, it now appears probable that at heavy pion masses
  there is no nucleon-nucleon bound state.}

\FullConference{The 11th International Workshop on Chiral Dynamics (CD2024)\\
 26-30 August 2024\\
Ruhr University Bochum, Germany\\}

\renewcommand{\logo}{\relax}

\renewcommand{\hookAfterAbstract}{%
  \par\bigskip
  \hfill DESY-25-027
}

\newcommand{\op}{\mathcal{O}}
\newcommand{\Cmat}{\mathbf{C}}

\begin{document}
\maketitle

\section{Introduction}

Lattice QCD calculations of multi-baryon systems have the ambition of
deriving few-baryon interactions, which form the basis of nuclear and
hypernuclear physics, starting from the Standard Model of particle
physics using just a handful of experimental inputs such as the masses
of the pion, kaon, and nucleon.

In addition to replacing experimental input for two- and three-baryon
interactions, lattice QCD calculations can also answer questions not
easily accessed. At unphysical quark masses, these calculations
provide information about the dependence of the deuteron binding
energy on the parameters of the Standard Model, which serves as input
for studies of the fine-tunedness of the universe and whether the
parameters could have been different during Big Bang
nucleosynthesis~\cite{Meyer_CD2024}. Unlike in experiment, systems
with strangeness are not especially challenging on the lattice, making
them likely the first place lattice calculations will have an impact
on phenomenology; it is hoped that precise nucleon-hyperon,
hyperon-hyperon, and eventually nucleon-nucleon-hyperon interactions
will help to resolve the hyperon puzzle in neutron
stars~\cite{Nogga_CD2024}.

Beyond scattering, lattice calculations can also include external
probes, gaining insight into the structure of light nuclei. The EMC
effect and quenching of axial charge revealed that interactions of a
probe with more than one nucleon can contribute significantly, making
it important to control these effects in particle physics experiments
that use nuclear targets such as dark matter direct detection and
neutrino experiments. Neutrinoless double beta decay, an inherently
two-nucleon process that violates lepton number and can occur if
neutrinos have a Majorana mass, is also a long-term
goal~\cite{deVries_CD2024}; it is further complicated by the need to
deal with two separate currents.

Lattice QCD calculations of baryon-baryon interactions are in a state
of change, transitioning from an earlier generation of exploratory
calculations to new calculations with a focus on understanding and
controlling systematic uncertainties, based in part on methods adapted
from the last decade of successful meson-meson calculations. From
these new calculations, it now appears clear that the earlier
calculations were afflicted by large uncontrolled systematic
uncertainties and that claims of a deeply bound deuteron and dineutron
at heavy pion mass are not correct.

In this review, I will focus on calculations that use the periodic box
as a tool, i.e.\ those based on finite-volume spectroscopy and
Lüscher's quantization condition~\cite{Luscher:1990ux,
  Rummukainen:1995vs, Briceno:2013lba,
  Briceno:2014oea}. Alternatively, the HAL~QCD method obtains
hadron-hadron potentials from equal-time Nambu-Bethe-Salpeter wave
functions. It has different sources of systematic uncertainty and has
never claimed a deeply bound nucleon-nucleon bound state. I will say
little about calculations using this method; the reader is referred to
the review by Sinya Aoki at this workshop~\cite{Aoki_CD2024}.

\section{Methods and challenges}

The standard approach for hadron-hadron interactions on the lattice
proceeds by the following steps:
\begin{enumerate}
\item Compute the low-lying finite-volume energy levels for various
  quantum numbers according to symmetries of the lattice theory:
  flavour, total momentum $\vec P$, and irreducible representation
  (irrep) $\Lambda$ of the little group containing lattice rotations
  and reflections that preserve $\vec P$. Since lattice calculations
  are done in cubic finite volumes of size $L$ with periodic boundary
  conditions, $\vec P$ is quantized in integer multiples of
  $2\pi/L$. In the continuum in infinite volume, the little group
  irrep is given by the $J^P$ quantum numbers, whereas in finite
  volume the little group is much smaller and multiple $J^P$ can mix
  in the same irrep.
\item Use finite-volume quantization conditions to constrain the
  scattering amplitude $T$. Each energy level $E$ implies one
  constraint on the infinite set of channels and partial wave
  amplitudes at the corresponding centre-of-mass energy
  $T_\Lambda(E_\text{cm})$ that contribute to $\Lambda$. In practice,
  one must neglect partial waves beyond the lowest few and it is
  necessary to employ a model describing the energy-dependence of
  $T$. The energy levels then serve as constraints on the parameters
  of the model.
\item By interpolating the model along the real energy axis or
  extrapolating into the complex plane, find poles corresponding to
  bound states and resonances. Using a variety of models, one can
  judge the robustness of the poles and estimate systematic
  uncertainty from the modelling.
\end{enumerate}
In addition, one must control the standard lattice systematics (which
also affect the HAL~QCD method): discretization effects coming from a
nonzero lattice spacing $a$, residual finite-volume effects, and
unphysical quark masses $m_q$. Ideally, the calculation is
extrapolated $a\to 0$, $L\to\infty$, and $m_q\to m_q^\text{phys}$ (or
calculated directly at $m_q^\text{phys}$). Here I will focus on the
first step --- computing the finite-volume spectrum --- and on
discretization effects, which may be more important than was expected.

\subsection{Finite-volume spectroscopy}

\subsubsection{Excited-state effects and signal-to-noise ratio}

Finite-volume spectroscopy starts with two-point correlation functions
of time-local interpolating operators $\op$ and $\op'$ with the
desired quantum numbers, separated in Euclidean time. Inserting a
complete set of states, we get
\begin{equation}
  \begin{aligned}
  C(t) \equiv \left\langle \op'(t) \op^\dagger(0) \right\rangle
  &= \sum_n e^{-E_n t} \langle\Omega|\op'|n\rangle
  \langle n|\op^\dagger|\Omega\rangle \\
    &\xrightarrow{t \gg (E_1-E_0)^{-1}} e^{-E_0 t}
  \langle\Omega|\op'|0\rangle
    \langle 0|\op^\dagger|\Omega\rangle,
  \end{aligned}
\end{equation}
where $|\Omega\rangle$ is the vacuum and $|n\rangle$ are the
finite-volume states of the specified quantum numbers, ordered by
increasing energy. As $t$ increases, Euclidean time evolution acts as
an exponential filter to remove excited states, and it is eventually
possible to extract the ground-state energy from the exponential decay
of $C(t)$. Furthermore, if $\op=\op'$ then every term in the sum is
positive and the effective energy
$E_\text{eff}(t)\equiv -\frac{d}{dt}\log C(t)$ will approach $E_0$
monotonically from above.

Complicating the matter is the well-known signal-to-noise
problem~\cite{Parisi:1983ae, Lepage:1989hd}. For a single-nucleon
correlation function, the statistical error decays asymptotically with
$\exp(-\frac{3}{2}m_\pi t)$, more slowly than the correlator
decays. This loss of signal makes it impossible in practice to use
very large $t$. One is reduced to looking for an acceptable
``window'', where the time separation is large enough to suppress
excited-state contributions but small enough that the signal is still
good.

\begin{figure}
  \centering
  \includegraphics[width=0.495\textwidth]{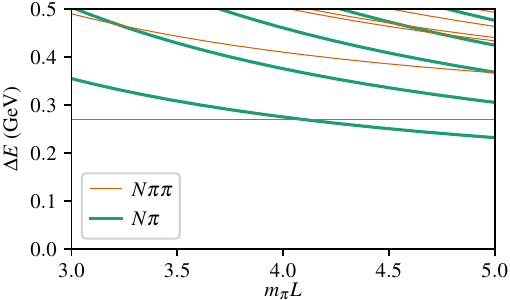}
  \includegraphics[width=0.495\textwidth]{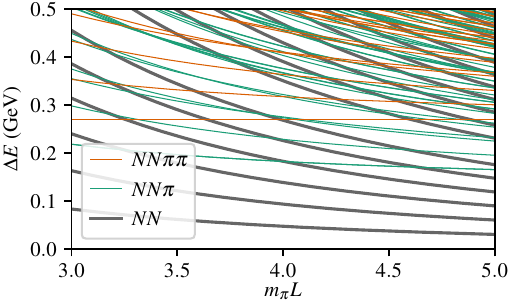}
  \caption{Energy gaps $\Delta E$ between ground state and
    noninteracting multiparticle energy levels, versus box size in
    units of $m_\pi^{-1}$: single-nucleon sector (left) and
    two-nucleon sector (right). Here the physical pion mass is used
    and systems with zero total momentum are shown.}
  \label{fig:dE_free}
\end{figure}

Figure~\ref{fig:dE_free} illustrates the approximate energy gaps
between the ground state and excited states, neglecting interactions
between hadrons, as functions of $L$. In an actual calculation, the
energy levels will be shifted due to those interactions as governed
(in the two-particle regime) by Lüscher's quantization condition. For
a single nucleon at rest, states with the finite-volume version of
$J^P=\frac{1}{2}^+$ quantum numbers can contribute. Here, the minimum
energy gap is roughly $2m_\pi$: one can excite two pions at rest or
one pion with a single unit of back-to-back momentum given to the pion
and the nucleon. For a typical calculation with $m_\pi L=4$, these two
excitations are nearly degenerate, and as $L\to\infty$, the lowest
energy gap becomes $m_\pi$.

For a two-nucleon system, the lowest energy gap is much smaller
because the system can be elastically excited, i.e.\ each nucleon can
receive one unit of momentum, yielding a gap of approximately
$4\pi^2/(m_NL^2)$. For the same typical calculation at the physical
pion mass, this corresponds to slightly more than $m_\pi/3$. Thus
Euclidean time evolution is only one sixth as effective at filtering
out excited states than in the single-nucleon case. In addition, the
signal-to-noise ratio decays asymptotically twice as fast. The
combination of these two problems is the main reason why two-nucleon
systems are so challenging on the lattice: there may be no window in
which the ground state has been reached but there is still a signal.

\subsubsection{Improved methods}

One approach to improve spectroscopy calculations is to suppress the
noise. This requires modifying the standard algorithmic strategy:
multilevel methods have been developed for this
purpose~\cite{Ce:2016idq, Ce:2016ajy, Giusti:2022xdh}, although they
have not yet been used in a large-scale calculation. Alternatively,
the filtering out of excited states can be made faster using
variational methods, as is now standard in multi-meson
spectroscopy. Rather than computing a single correlation function, one
computes an $N\times N$ matrix
$\Cmat_{ij}(t) \equiv \langle \op_i^{\vphantom{\dagger}}(t)
\op_j^\dagger(0) \rangle$ using $N$ different interpolating operators
$\op_i$~\cite{Michael:1985ne, Luscher:1990ck}. By solving a
generalized eigenvalue problem (GEVP)
\begin{equation}
  \Cmat(t) v_n = \lambda_n \Cmat(t_0) v_n,
\end{equation}
for each of the lowest $N$ states one effectively finds an optimal
linear combination $\tilde \op_n \equiv (v_n^\dagger)_i \op_i$ for
isolating it. The energy of state $n$ can also be estimated with a
systematic error that decays asymptotically as
$e^{-(E_N-E_n)t}$~\cite{Blossier:2009kd}. For the ground state, this
removes the contributions from the first $N-1$ excited states,
yielding a much faster filtering by Euclidean time evolution.

\subsubsection{Interpolating operators}

Whether one uses a single correlator or a matrix, the starting point
is interpolating operators (interpolators). Typically these are
constructed using ``smeared'' quark fields with a spatial extent
similar to the hadrons of interest. Two kinds have seen widespread use
for studying two-baryon systems. The first is local ``hexaquark''
operators of form
\begin{equation}
  \op_H(t) = \sum_{\vec x} e^{-i\vec P\cdot \vec x} (qqqqqq)(t,\vec x),
\end{equation}
i.e.\ with all six quarks at the same point. (Here, details of
flavour, colour, and spin have been omitted.) These are similar to
states like the $H$~dibaryon in the quark model. The second is bilocal
``baryon-baryon'' operators,
\begin{equation}
  \op_{BB}(t) = \sum_{\vec x,\vec y} e^{-i \vec p_1\cdot \vec x}
  e^{-i(\vec P-\vec p_1)\cdot \vec y} (qqq)(t,\vec x) (qqq)(t,\vec y),
\end{equation}
which factorize into the product of two ``baryon'' operators of
definite momentum and thus have the structure of a noninteracting
two-baryon state. Varying the single-baryon momentum $\vec p_1$ allows
the construction of many different operators with the same total
momentum $\vec P$.

Given a choice of interpolating operators, computing their two-point
correlation function is a nontrivial task. Since the fermionic path
integral is done analytically, one must evaluate Wick contractions of
quark propagators $D^{-1}(x,y)$ on each gauge background. Efficient
solvers have been developed for the linear system $D\psi=\phi$, which
yield $D^{-1}$ applied to a specified source $\phi$. Many calculations
in the past have used simple smeared point sources, which imply that
all smeared quarks in the creation operator $\op^\dagger$ lie at the
same point, allowing only hexaquark interpolators. This was often
paired with a baryon-baryon annihilation operator $\op'$. Newer
calculations use more sophisticated techniques such as
distillation~\cite{HadronSpectrum:2009krc, Morningstar:2011ka} or
sparsening~\cite{Detmold:2019fbk}, allowing a greater variety of
interpolators and the use of variational methods.

\subsection{Finite-volume quantization conditions}

When a bound state is present, the finite-volume ground state will
approach the bound-state mass $m_\text{bound}$ with finite-volume
corrections that are exponentially suppressed with $e^{-\gamma L}$,
where $\gamma$ is the binding momentum with respect to the nearest
two-particle scattering threshold for particles of mass
$m_\text{scatt}$, i.e.\ 
$m_\text{bound} = 2\sqrt{m_\text{scatt}^2-\gamma^2}$. For shallow
bound states such as the deuteron, $\gamma$ can be small, producing
slowly decaying finite-volume effects. In these cases, finite-volume
quantization conditions can serve as a workaround: the finite-volume
energies constrain the two-particle scattering amplitude, both above
and below the bound-state mass. By interpolating, the bound-state pole
can be found with much smaller finite-volume effects.

Following the initial work by Lüscher~\cite{Luscher:1990ux},
two-particle finite-volume quantization conditions have been
generalized to account for nonzero momentum, spin, and coupled
channels~\cite{Rummukainen:1995vs, Briceno:2013lba,
  Briceno:2014oea}. Following Ref.~\cite{Morningstar:2017spu}, it can
be written as $\det(\tilde K^{-1}-B)=0$, where $\tilde K$ is the
reduced $K$-matrix, related algebraically to the infinite-volume
scattering amplitude; $B$ is the finite-volume matrix that depends on
$\vec P$, $L$, and little-group irrep; and the determinant runs over
channels and partial waves. Given an ansatz for $\tilde K$ as a
function of energy, its solutions are the predicted finite-volume
energy levels. Truncated to a single channel and partial wave, it
becomes a simple equation giving the phase shift $\delta(E_\text{cm})$
at the centre-of-mass energy corresponding to each finite-volume
energy level.

It has been known for a long time that finite-volume quantization
conditions break down above the lowest untreated threshold, which
typically corresponds to three particles. However, it was noted in
Ref.~\cite{Green:2021qol} that a breakdown also occurs somewhere below
the lowest two-particle threshold, due to left-hand cuts. This is
particularly relevant for baryon-baryon systems, in which pion
exchange causes left-hand cuts to occur close to the lowest
threshold. With several new proposals on how to deal with this
issue~\cite{Meng:2021uhz, Meng:2023bmz, Raposo:2023oru,
  Hansen:2024ffk, Dawid:2024dgy, Bubna:2024izx, Dawid:2024oey}, this
is an ongoing area of research and the reader is referred to the talks
on this subject presented at this workshop~\cite{Hansen_CD2024,
  Rusetsky_CD2024, Meng_CD2024, Dawid_CD2024,
  RomeroLopez_CD2024}. Another direction being investigated is how to
include lattice artifacts that break $O(4)$
symmetry~\cite{Hansen:2024cai}.

\section{Old and new calculations}

\begin{figure}
  \centering
  \includegraphics[width=0.49\textwidth]{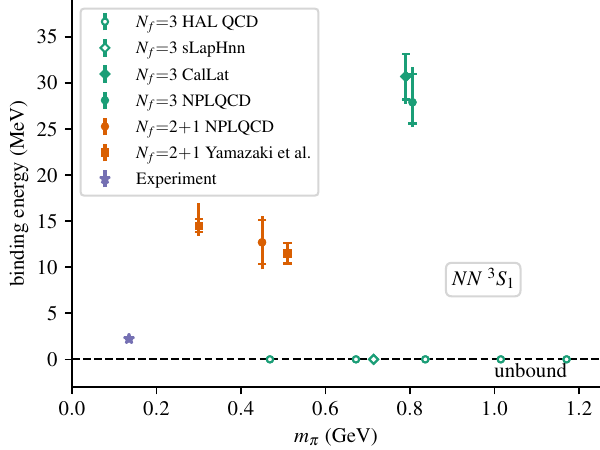}
  \includegraphics[width=0.49\textwidth]{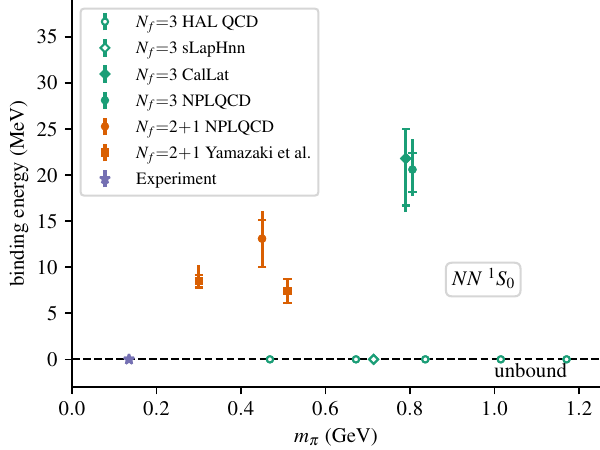}
  \caption{Binding energy of the deuteron (left) and dineutron (right)
    versus pion mass~\cite{Inoue:2011ai, Horz:2020zvv,
      Berkowitz:2015eaa, Wagman:2017tmp, NPLQCD:2020lxg,
      Yamazaki:2012hi, Yamazaki:2015asa}. Points on the horizontal
    dashed line denote no bound state. Equal light and strange quark
    masses are denoted in green, whereas unequal masses (with
    near-physical $m_s$) are denoted in orange.}
  \label{fig:old_NN}
\end{figure}

There has been a decade-long controversy over whether nucleon-nucleon
bound states exist for QCD with heavier-than-physical quark masses;
see Fig.~\ref{fig:old_NN}. Some calculations reported that the
deuteron becomes more deeply bound as the pion mass is increased, and
that a second ``dineutron'' (isospin one, spin zero) bound state
appears~\cite{Yamazaki:2015asa, Yamazaki:2012hi, Berkowitz:2015eaa,
  NPLQCD:2011naw, NPLQCD:2012mex, Orginos:2015aya, Wagman:2017tmp,
  NPLQCD:2020lxg}. However, first HAL~QCD and then others reported the
opposite result: that there is no $NN$ bound state~\cite{Inoue:2011ai,
  Horz:2020zvv}. More recent calculations for which preliminary
results were reported~\cite{Green:2022rjj} or where conclusions about
the presence of bound states were not
reached~\cite{Amarasinghe:2021lqa, Detmold:2024iwz} are also
consistent with there being no bound state.

The presence or absence of nucleon-nucleon bound states is one of the
simplest questions about nuclear physics that can be asked of lattice
QCD. This long-standing disagreement is a significant problem that
needs to be resolved before lattice calculations of other two-baryon
observables can be taken seriously. Fortunately, a new generation of
calculations promises to better control systematic uncertainties and
resolve the controversy.

What explains the disagreement in Fig.~\ref{fig:old_NN}? One possible
source is discretization effects: except for recent calculations in
Refs.~\cite{Green:2021qol, Green:2021sxb, Green:2022rjj,
  Inoue:2024osj, Perry_Lat2024}, no baryon-baryon study used more than
one lattice spacing $a$. Thus, the missing $a\to 0$ extrapolation is a
completely uncontrolled source of systematic error. Another source is
excited states contaminating the estimate of the ground-state
energy. Calculations that obtained bound states used point-source
methods and asymmetric correlation functions of form
$\langle \op_{BB}(t) \op^\dagger_H(0) \rangle$, whereas calculations
yielding no bound state used either the HAL~QCD method (with very
different systematics) or variational methods~\cite{Horz:2020zvv,
  Green:2022rjj, Amarasinghe:2021lqa, Detmold:2024iwz}.

\begin{figure}
  \centering
  \includegraphics[width=0.49\textwidth]{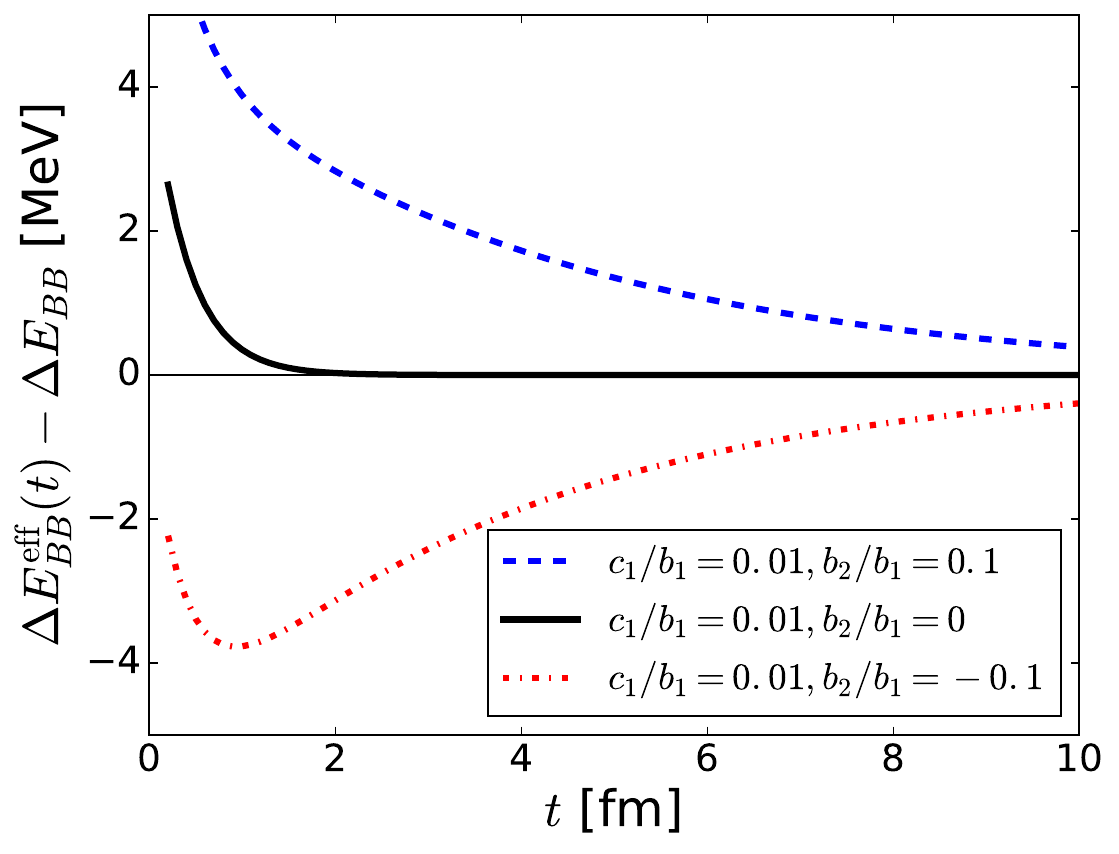}
  \includegraphics[width=0.49\textwidth]{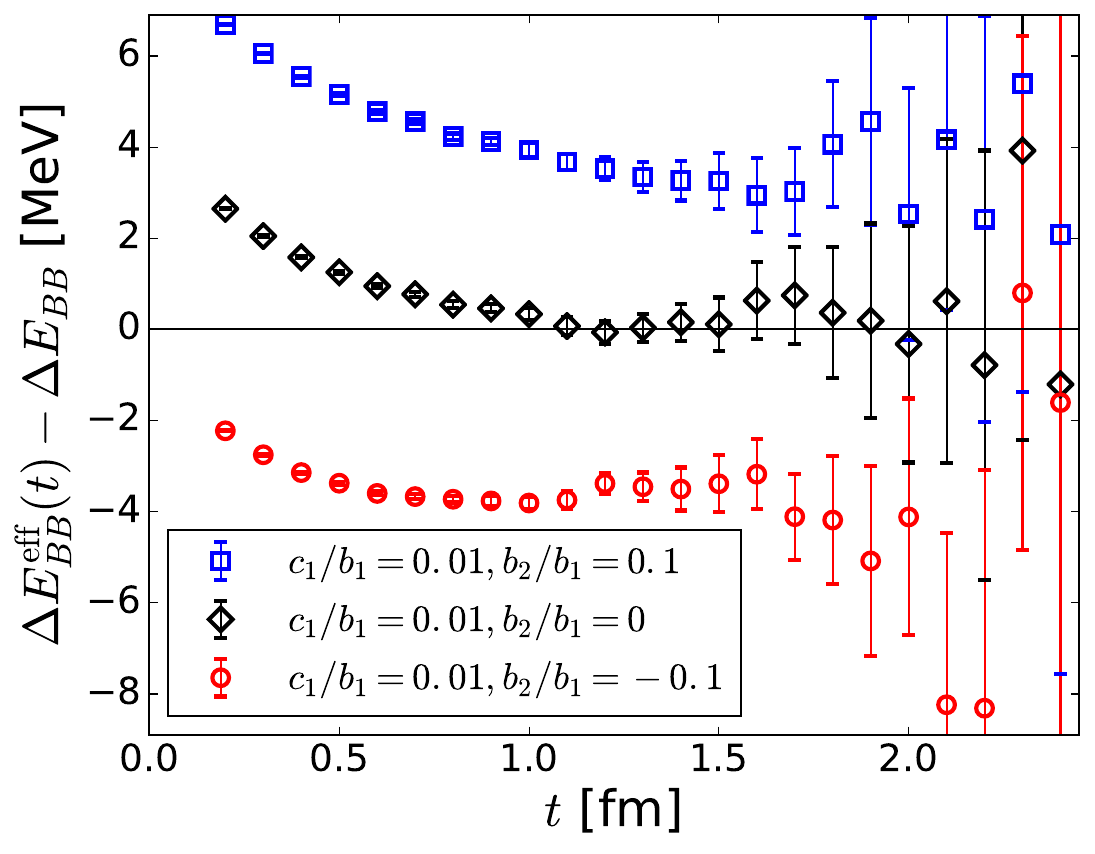}
  \caption{Mock data illustrating a scenario where a ``fake plateau''
    appears: effective energy
    $E_\text{eff}(t)\equiv -\frac{d}{dt}\log C(t)$ for correlator
    $C(t)=b_1 + b_2 e^{-\delta E_\text{el} t} + c_1 e^{-\delta
      E_\text{inel} t}$ with small ``elastic'' energy gap
    $\delta E_\text{el}=50$~MeV and large ``inelastic'' energy gap
    $\delta E_\text{inel}=500$~MeV. The blue dashed curve and squares,
    black solid curve and diamonds, and red dash-dotted curve and
    circles correspond to positive, zero, and negative
    elastic-excitation amplitude $b_2$, respectively. All data
    asymptotically approach zero as $t\to\infty$. Left: curves without
    error, for a wide range of $t$. Right: points with added noise to
    simulate lattice data. This figure is reproduced from
    Ref.~\cite{Iritani:2016jie} under the Creative Commons Attribution
    License.}
  \label{fig:HAL_fake}
\end{figure}

The question of how a spectroscopy calculation can go wrong has been
studied by HAL~QCD~\cite{Iritani:2016jie}. If variational methods are
not used, then the very small energy gap $\Delta E$ between the ground
state and lowest-lying elastic excitation controls the asymptotic
approach to the energy plateau. The need for $t\gg (\Delta E)^{-1}$
implies that Euclidean time separations $t$ of up to 10~fm are needed
to isolate the ground-state energy. Furthermore, an asymmetric
correlation function allows excited-state contributions of either
sign, further confounding the plateau. Given that the noise starts to
grow rapidly for $t\gtrsim 1.5$~fm, this can cause ``fake plateaus''
to appear; see Fig.~\ref{fig:HAL_fake}. In particular, noise
transforms the case with a local minimum into the best-looking
plateau.

\begin{SCfigure}
  \centering
  \includegraphics[width=0.5\textwidth]{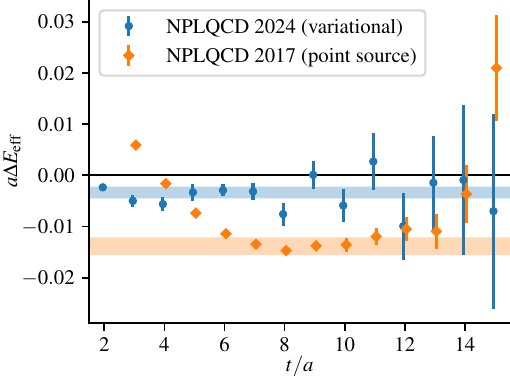}
  \caption{Effective energy difference for the rest-frame ground-state
    $NN$ $^1S_0$ energy: point-source and variational methods applied
    to the same $24^3$ ensemble. Data are extracted from
    Refs.~\cite{Wagman:2017tmp, Detmold:2024iwz}.}
  \label{fig:NPLQCD_comparison}
\end{SCfigure}

This effect can be seen in two sets of calculations done by NPLQCD on
the same lattice ensembles at a heavy SU(3)-symmetric point
($m_\pi=m_K=m_\eta\approx 800$~MeV). The earlier
work~\cite{NPLQCD:2012mex, Wagman:2017tmp}, which used point sources
and asymmetric correlators, found a dineutron ($NN$ $^1S_0$) bound
state with binding energy $B_{nn}\approx 21$~MeV. The newer
work~\cite{Amarasinghe:2021lqa, Detmold:2024iwz}, which used
variational techniques with a variety of local and bilocal
interpolating operators, is consistent with no bound state. As shown
in Fig.~\ref{fig:NPLQCD_comparison}, the two calculations produced
quite incompatible energy plateaus. Since variational methods yield
better control over low-lying excited states, the newer calculation
has a more trustworthy plateau. For further discussion on this issue,
see Refs.~\cite{Iritani:2018vfn, INT_perspective}.

Over the last few years, an increasing number of independent
collaborations have applied variational methods to baryon-baryon
systems~\cite{Francis:2018qch, Horz:2020zvv, Amarasinghe:2021lqa,
  Green:2021qol, Green:2021sxb, Green:2022rjj, Detmold:2024iwz,
  Geng:2024dpk, Wang_Lat2024, Perry_Lat2024}. A largely consistent
picture is starting to emerge: there is no nucleon-nucleon bound state
at heavy pion masses.

\section{Calculations at light SU(3)-symmetric point}

In this section I will report some past and ongoing calculations that
I have done together with the ``Mainz group'' and the broader Baryon
Scattering (BaSc) collaboration. These have been done using ensembles
from CLS~\cite{Bruno:2014jqa} at an SU(3)-symmetric point where the
up, down, and strange quarks are degenerate with mass set to the
average of their masses in nature, corresponding to
$m_\pi=m_K=m_\eta\approx 420$~MeV. We employed variational methods to
study multiple low-lying energy levels and eight lattice ensembles
covering a range of different box sizes $L$ and lattice spacings $a$.

\subsection{Nucleon-nucleon $^1P_1$ and $^1F_3$}

\begin{figure}
  \centering
  \includegraphics[width=\textwidth]{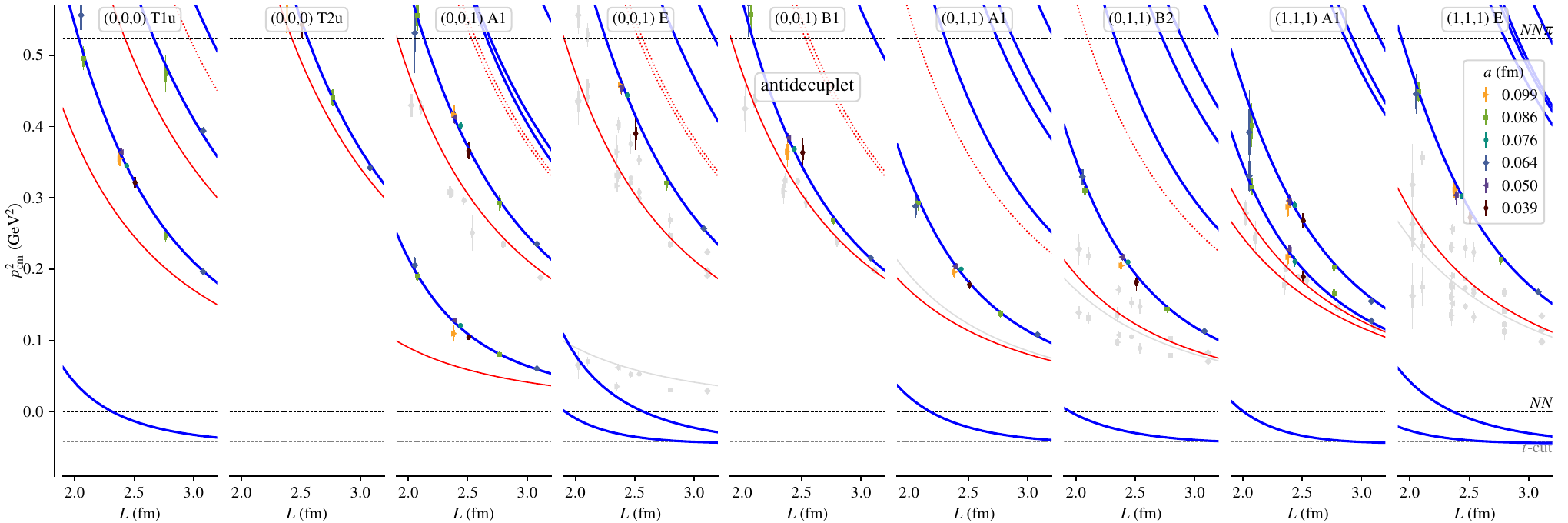}
  \caption{(preliminary) Finite-volume energy levels relevant for $NN$
    spin-zero odd partial waves, versus box size~\cite{Green:2022rjj,
      inprep_NN}. Energy has been transformed to centre-of-mass
    momentum squared, $E_\text{cm}^2=4(m_N^2+p^2_\text{cm})$. The
    different panels correspond to different total momentum, indicated
    by $\vec PL/(2\pi)$, and different little-group irrep. Horizontal
    lines indicate thresholds and the left-hand cut, red curves show
    noninteracting levels, points show lattice-computed energy levels,
    and blue curves show the best fit based on applying finite-volume
    quantization to models of the $^1P_1$ and $^1F_3$ phase
    shifts. Different coloured points indicate different lattice
    spacings as given in the legend. Gray points and curves indicate
    spin-one energy levels with the same finite-volume quantum
    numbers: the lattice levels have been identified via their
    dominant coupling to spin-one interpolators.}
  \label{fig:I0_spin0}
\end{figure}

Flavour symmetries are unbroken on the lattice and the two-particle
quantization condition does not mix different spins, allowing us to
analyze four different subsets of nucleon-nucleon energy levels
independently: combining spin zero or one with isospin zero or
one. The simplest of these to analyze is isospin-zero spin-zero, which
contains odd partial waves $^1P_1$, $^1F_3$, $^1H_5$, etc. The
corresponding energy levels are shown in Fig.~\ref{fig:I0_spin0}. In
all cases, the lattice energy is shifted upward relative to the
noninteracting one, indicating a repulsive interaction.

The lattice spectrum can be fitted using a chiral EFT-inspired model
for the $^1P_1$ and $^1F_3$ phase shifts combined with finite-volume
quantization. I include the one-pion-exchange potential and one
contact term in each partial wave, regularized following
Ref.~\cite{Reinert:2017usi} with cutoff $\Lambda=1.5m_\pi$, and solve
the Lippmann-Schwinger equation to obtain the phase shifts. With just
three fit parameters ($g_{\pi NN}$ and the coefficients of the two
contact terms) and assuming no discretization effects, this model
provides a good description of the spectrum and a reasonable-valued
coupling constant $g_{\pi NN}\approx 13$. However, in a sign of the
breakdown of the quantization condition, there are also spurious
solutions that converge toward the onset of the left-hand cut as
$L\to\infty$. Here they can be easily skipped over, since the spectra
start well above threshold; however, spurious solutions are
potentially problematic for analyzing $S$ waves, which are constrained
by near-threshold energy levels.

\begin{figure}
  \centering
  \includegraphics[width=0.49\textwidth]{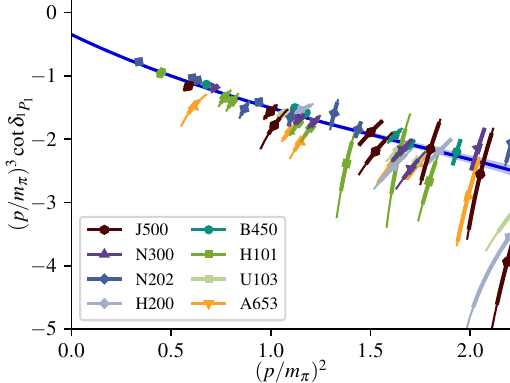}
  \includegraphics[width=0.49\textwidth]{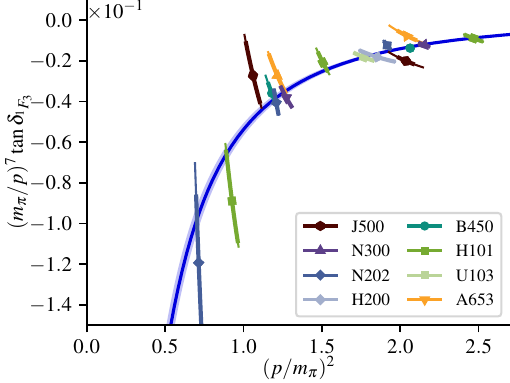}
  \caption{(preliminary) Nucleon-nucleon spin-zero odd-partial-wave
    phase shifts: $p^3\cot\delta_{^1P_1}$ (left) and
    $(p^7\cot\delta_{^1F_3})^{-1}$ (right) versus $p^2$, in units of
    the pion mass~\cite{Green:2022rjj, inprep_NN}. Blue curves show
    the best fit described in the text. Points in the left plot show
    the $^1P_1$ phase shift determined from each lattice energy level,
    assuming the $^1F_3$ phase shift is given by the best fit, and
    vice versa for the right plot. The labels are ensemble names from
    CLS (see the inset of Fig.~\ref{fig:H_spectrum}); darker and more
    purple colours indicate finer lattice spacings and the two pale
    colours indicate small volumes.}
  \label{fig:1P1_1F3}
\end{figure}

The corresponding phase shifts are shown in Fig.~\ref{fig:1P1_1F3}. In
a highly nontrivial check of the finite-volume spectrum, all of the
phase-shift points from different volumes, moving frames, and irreps
lie on a single curve, suggesting that the uncertainties of the
spectrum are under reasonable control. Furthermore, the different
lattice spacings agree, indicating that discretization effects are
small.

\subsection{$H$ dibaryon}

Conjectured decades ago based on an MIT bag model
calculation~\cite{Jaffe:1976yi}, the $H$~dibaryon is a hypothetical
SU(3) flavour singlet scalar that would be a $\Lambda\Lambda$ bound
state. Past lattice calculations~\cite{Inoue:2010es, Inoue:2011ai,
  HALQCD:2019wsz, NPLQCD:2010ocs, Beane:2011zpa, NPLQCD:2011naw,
  NPLQCD:2012mex} have also found bound states for heavy quark masses,
with binding energies of up to 75~MeV. This includes calculations by
HAL~QCD, who have consistently found no nucleon-nucleon bound
state. However, at similar quark masses corresponding to
$m_\pi\approx 800$~MeV, the binding energy reported by HAL~QCD is
about half as big as the one reported by NPLQCD.

\begin{figure}
  \centering
  \includegraphics[width=\textwidth]{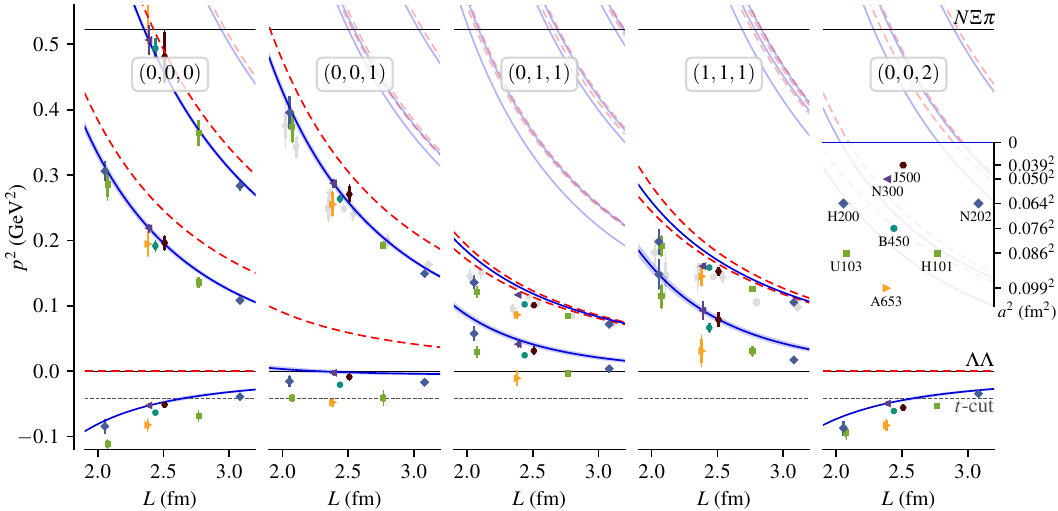}
  \caption{Finite-volume energy levels relevant for the SU(3) singlet
    baryon-baryon $^1S_0$ partial wave; see the caption of
    Fig.~\ref{fig:I0_spin0}. The trivial irrep ($A_{1g}$ or $A_1$) is
    shown for each moving frame. The blue curves show the fit model
    evaluated in the continuum limit. This figure is reproduced from
    Ref.~\cite{Green:2021qol} under the Creative Commons Attribution
    License.}
  \label{fig:H_spectrum}
\end{figure}

Our finite-volume spectra relevant for the $H$~dibaryon at the light
SU(3)-symmetric point are shown in Fig.~\ref{fig:H_spectrum}. There
are two notable features. First, there is a strong dependence on the
lattice spacing: the difference between noninteracting and lattice
energies can vary by a factor of almost two between the finest and
coarsest lattices. This was a surprising result, as it had been
expected that discretization effects would mostly cancel in this
difference. Second, the lowest energy levels lie on top of the
left-hand cut, making them unusable in Lüscher's finite-volume
quantization condition.

In Ref.~\cite{Green:2021qol}, we fitted these data by describing
$p\cot\delta(p)$ with polynomials in $p^2$, allowing the coefficients
to depend on the lattice spacing. The levels on the left-hand cut were
omitted from the analysis. In the continuum limit, we found a binding
energy of roughly 5~MeV, whereas at nonzero lattice spacing this was
up to seven times larger. This difference is comparable in size to the
spread between past studies of the $H$~dibaryon, which each used just
a single lattice spacing.

For two-nucleon energy levels, preliminary results in irreps that
contain $^1S_0$ and $^3S_1$--$^3D_1$ partial waves also show
significant dependence on the lattice spacing~\cite{Green:2021sxb,
  Green:2022rjj}, in contrast with the spin-zero odd partial waves
shown in the previous subsection. Recently, significant lattice
artifacts for the $H$~dibaryon binding energy have also been found by
HAL~QCD~\cite{Inoue:2024osj} (with otherwise very different
systematics) and NPLQCD~\cite{Perry_Lat2024}. Thus, the need for fine
lattice spacings and good control over the continuum limit adds to the
challenges in studying two-baryon systems on the lattice.

\begin{SCfigure}
  \centering
  \includegraphics[width=0.6\textwidth]{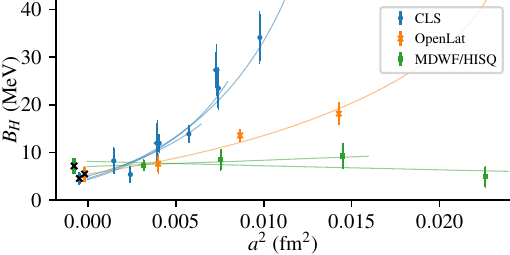}
  \caption{(preliminary) Binding energy of the $H$ dibaryon versus
    squared lattice spacing~\cite{Green_Lat2024, inprep_Hdib}. Three
    independent continuum extrapolations with different lattice
    actions yield consistent binding energies in the $a\to 0$
    continuum limit (black crosses).
  }
  \label{fig:dE_vs_a2}
\end{SCfigure}

Together with the broader BaSc collaboration, I have been studying the
origin of these discretization effects by repeating the study of the
$H$~dibaryon using different lattice formulations of
QCD~\cite{Green_Lat2024, inprep_Hdib}. We supplemented the existing
calculation on CLS ensembles using clover fermions from
Ref.~\cite{Green:2021qol} with two other actions with the same quark
masses. First, we used three ensembles with a modified clover
term~\cite{Francis:2019muy} from
OpenLat~\cite{Francis:2023gcm}. Second, we generated four
staggered-fermion ensembles with the HISQ action~\cite{Follana:2006rc,
  MILC:2010pul} (including a physical-mass charm quark) and used
Möbius domain wall valence fermions in a mixed-action
setup~\cite{Berkowitz:2017opd}. As shown in Fig.~\ref{fig:dE_vs_a2},
these other actions produce a much weaker dependence on the lattice
spacing, which might help future calculations control discretization
effects at a lower computational cost.

\section{Summary and outlook}

Following increased scrutiny, it now appears probable that past
exploratory lattice QCD calculations of two-baryon systems were
affected by two sources of large uncontrolled systematic uncertainty:
the ground-state energy was not adequately isolated and discretization
effects were neglected. In particular, there is no nucleon-nucleon
bound state at significantly-heavier-than-physical pion masses.

Moving forward, the community needs to maintain its focus on
controlling systematic effects. A better understanding of lattice
artifacts, their origin, and how to model them, would be quite
valuable. Applying modified quantization conditions that account for
the left-hand cut will help to control and understand the residual
finite-volume effects. In general, it will be important to have more
detailed cross-checks between different collaborations including
HAL~QCD (with its different methodology): rather than simply agreeing
on the absence of a bound state, we should quantitatively compare
scattering lengths, effective ranges, phase shifts, and mixing angles.

Finally, we need to push to lighter quark masses, towards the physical
point. As the signal-to-noise problem becomes more severe and the
energy splittings become smaller, new techniques may be needed to make
the calculations feasible.

\acknowledgments

I thank my colleagues in the BaSc collaboration for valuable
discussions.
Calculations for these projects used resources
provided by the John von Neumann Institute for Computing and Gauss
Centre for Supercomputing e.V.\ (\url{www.gauss-centre.eu}) on
JUQUEEN~\cite{juqueen}, JURECA~\cite{jureca}, and JUWELS~\cite{juwels}
at Jülich Supercomputing Centre;
on Summit at the Oak Ridge Leadership Computing Facility at the Oak
Ridge National Laboratory, which is supported by the Office of Science
of the U.S. Department of Energy under Contract No. DE-AC05-00OR22725;
provided by the Lawrence Livermore National Laboratory (LLNL)
Multiprogrammatic and Institutional Computing program for Grand
Challenge allocations on Lassen at LLNL;
and of the National Energy Research Scientific Computing Center
(NERSC), a Department of Energy Office of Science User Facility using
NERSC awards NP-ERCAP0033436 and NP-ERCAP0027666.
This work used the software packages GLU~\cite{GLU},
QDP++~\cite{Edwards:2004sx}, PRIMME~\cite{PRIMME},
openQCD~\cite{openQCD}, QUDA~\cite{Clark:2009wm, Babich:2011np,
  Clark:2016rdz}, opt\_einsum~\cite{opt_einsum},
SigMonD~\cite{sigmond}, TwoHadronsInBox~\cite{Morningstar:2017spu},
NumPy~\cite{numpy}, SciPy~\cite{scipy}, and
Matplotlib~\cite{Hunter:2007}.
We are grateful to our colleagues within CLS and OpenLat for sharing
ensembles.

\bibliographystyle{JHEP_mod}
\bibliography{two-baryon.bib}

\end{document}